\documentclass{article}

\usepackage[nonatbib,final]{neurips_2021}
\usepackage[utf8]{inputenc} 
\usepackage[T1]{fontenc}    
\usepackage{hyperref}       
\usepackage{url}            
\usepackage{booktabs}       
\usepackage{amsfonts}       
\usepackage{nicefrac}       
\usepackage{microtype}      
\usepackage{xcolor}         

\usepackage{algorithm}
\usepackage{algpseudocode}
\usepackage{graphicx}
\usepackage[numbers,sort&compress]{natbib}
\usepackage{caption}
\usepackage{cleveref}
\usepackage[final]{pdfpages}

\title{Unsupervised Noise Adaptive Speech Enhancement by Discriminator-Constrained Optimal Transport}

\author{
  Hsin-Yi~Lin\\
  The Cooperative Institute for Research in Environmental Sciences (CIRES)\\
  University of Colorado, Boulder, CO, 80309, USA\\
  NOAA Physical Sciences Laboratory, Boulder, CO, 80305, USA\\
  \texttt{hylin@colorado.edu} \\
  \And
  Huan-Hsin~Tseng\\
  Research Center for Information Technology Innovation\\
  Academia Sinica, Taiwan\\
  \texttt{htseng@citi.sinica.edu.tw} \\
   \And
  Xugang~Lu\\
  National Institute of Information and Communications Technology (NICT), Japan\\
  \texttt{xugang.lu@nict.go.jp} \\
   \AND
  Yu~Tsao \\
  Research Center for Information Technology Innovation\\
  Academia Sinica, Taiwan\\
  \texttt{yu.tsao@citi.sinica.edu.tw} \\
}

\begin{document}
\maketitle

\begin{abstract}
This paper presents a novel discriminator-constrained optimal transport network (DOTN) that performs unsupervised domain adaptation for speech enhancement (SE), which is an essential regression task in speech processing. The DOTN aims to estimate clean references of noisy speech in a target domain, by exploiting the knowledge available from the source domain. The domain shift between training and testing data has been reported to be an obstacle to learning problems in diverse fields. Although rich literature exists on unsupervised domain adaptation for classification, the methods proposed, especially in regressions, remain scarce and often depend on additional information regarding the input data. The proposed DOTN approach tactically fuses the optimal transport (OT) theory from mathematical analysis with generative adversarial frameworks, to help evaluate continuous labels in the target domain. The experimental results on two SE tasks demonstrate that by extending the classical OT formulation, our proposed DOTN outperforms previous adversarial domain adaptation frameworks in a purely unsupervised manner.
\end{abstract}

\section{Introduction}
The goal of speech enhancement (SE) is to convert low-quality speech signals to ones with improved quality and intelligibility. SE serves as an important regression task in the speech-processing field and has been widely used for a pre-processor in speech-related applications, such as speech coding~\cite{li2011comparative}, automatic speech recognition (ASR)~\cite{li2015robust}, speaker recognition~\cite{michelsanti2017conditional}, and assistive hearing devices~\cite{wang2017deep, lai2016deep}. Recent advances in machine learning have made significant progress to the SE technology. Generally, learning-based SE approaches estimate a transformation to characterize the mapping function from noisy to clean speech signals in the training phase~\cite{wang2018supervised}. The estimated transformation converts noisy speech signals to generate clean-like signals in the testing phase. Various neural network models have been used to characterize  noisy-to-clean transformations. Well-known examples of such models include fully connected neural network~\cite{xu2014regression}, deep denoising autoencoder~\cite{lu2013speech}, convolutional neural network~\cite{fu2016snr}, long-short-term memory~\cite{chen2015speech}, and Transformer~\cite{koizumi2020speech}. To effectively handle diverse noisy conditions, we usually prepare a considerable amount of training data that cover various noise types to train SE models. However in real-application scenarios, the noise types in the testing data may not always be involved in the training set. Consequently, the noisy-to-clean transformation learned from the training data cannot be suitably applied to handle the testing noise, resulting in limited enhancement performance. This training-testing mismatch is generally called a domain mismatch issue for SE. An effective solution is required to perform domain adaptation to adjust the SE models with formulating a precise noisy-to-clean transformation that matches the testing conditions. Most existing domain adaptation methods rely on at least one of the following adaptation mechanisms: aligning domain-invariant features~\cite{tzeng2014deep, sun2016deep, morerio2017minimal} and adversarial training, where a discriminator is introduced during training as a domain classifier~\cite{ganin2015unsupervised, ganin2016domain,tzeng2017adversarial}.

This study aims to solve the unsupervised domain adaptation problem for SE by introducing optimal transport (OT). In particular, we consider the circumstance where SE is tested on a target domain with completely unlabeled data, and only labeled data from the source domain is available for reference. Generally speaking, OT theory compares two (probability) distributions and considers all possible transportation plans in between to find one with a minimal displacement cost. The concept of OT can be applied to minimize domain mismatch and consequently achieve unsupervised domain adaptation. Even with the mathematical characteristics offered by OT, obstacles to excellent SE performance persist due to the complex structure possessed by human speech. To further overcome the obstacles, another concept from Generative Adversarial Network (GAN) is integrated to assist attaining sophisticated SE domain adaptation. Although an existing domain transition technique ``domain adversarial training'' and our proposal share similarity in names, the fundamental constructions are substantially different. A key element in our method lies in a discriminator utilized to examine speech output characteristics, instead of a \textit{domain classifier}. More precisely, the discriminator in our method was employed to govern the output speech quality by learning the probability distribution of the source labels. This novel approach was designed especially for the unsupervised SE domain adaptation to exhibit excellent performance, which was verified on the VoiceBank and TIMIT datasets.

\paragraph{Contributions} 
We proposed a novel method designed specifically for unsupervised domain adaptation in a regression setting. This area of study still has very limited results; moreover, the existing methods often require additional classification of source domains or may not yet be supported by strong regression applications. Conversely, our approach does not require any additional input information other than the source samples, source labels, and target samples. Our approach was applied to two standardized SE tasks, namely VoiceBank-DEMAND and TIMIT, and  achieved superior adaptation performance in terms of both Perceptual Evaluation of Speech Quality (PESQ) and Short-Time Objective Intelligibility (STOI) scores. Furthermore, owing to the simple input requirements, we can easily investigate the effect of target sample complexity on our method by increasing the number of noise types allowed in the target domain, which has not been reported by previous literature to the best of our knowledge.

\section{Related work}
\paragraph{Adversarial domain adaptation:}
The main objective of Domain Adversarial Training (DAT) is to train a deep model (from the source domain) capable of adapting to other similar domains by leveraging a considerable amount of unlabeled data from the target domain~\cite{ganin2015unsupervised, ganin_book}. A conventional DAT system consists of three parts, deep feature extractor, label predictor, and domain classifier. By using a gradient reversal layer, the extracted deep features are discriminative for the main learning task and invariant with shifts between the source and target domains. The DAT approach has been applied and confirmed to effectively compensate for the mismatch of source (training time) and target (testing time) conditions in numerous tasks, such as speech signal processing~\cite{liao2018noise, serdyuk2016invariant}, image processing~\cite{ganin2015unsupervised, tang2020srda}, and wearable sensor signal processing~\cite{liu2020domain}. A later development in Multisource Domain Adversarial Networks (MDAN)~\cite{zhao2018adversarial} extended the original DAT to lift the constraint of single domain transition, utilizing  multiple domain classifiers to extract discriminative deep features for the main learning task while being invariant to multiple domain shifts~\cite{michelsanti2017conditional, richard2020unsupervised}.

\paragraph{Optimal transport for domain adaptation:}
Hitherto, OT ~\cite{villani2008optimal, peyre2019computational} has been utilized to domain adaptation~\cite{courty2014domain, courty2016optimal} with related analytical results~\cite{redko2017theoretical}. Furthermore, the concepts of OT have proved itself even more useful under a joint distribution framework~\cite{courty2017joint, damodaran2018deepjdot}. More recently, to improve the sensitivity of OT to outliers, Robust OT was proposed~\cite{balaji2020robust}. Moreover, a method for combining the notion of adversarial domain adaptation with OT and margin separation has been proposed~\cite{dhouib2020margin}. However, almost all experiments performed  in these studies were classification problems, unlike the SE task that is the focus of our study.

\paragraph{Domain adaptation in speech enhancement:}
Existing domain adaptation in SE approaches can be divided into two categories: supervised and unsupervised. For supervised domain adaptation, paired noisy and clean speech signals for the testing conditions are available to adjust the parameters in the SE models. In~\cite{xu2014cross, wang2015transfer}, transfer-learning-based approaches have been proposed to adapt SE models to alleviate corpus mismatches. To combat the catastrophic forgetting issue, Lee et al., proposed a SERIL algorithm that combines curvature-based regularization and path optimization augmenting strategies when preforming domain adaptation on SE models~\cite{lee2020seril}. Conversely, for unsupervised domain adaptation, only noisy speech signals are provided, and the corresponding clean counterparts are not accessible. Generally, unsupervised domain adaptation has good applicability to real-world scenarios. In~\cite{wang2020cross}, unsupervised domain adaptation for SE was performed by 
minimizing the Kullback-Leibler divergence between posterior probabilities produced by teacher and student senone classifiers without paired noisy-clean adaptation data. In~\cite{liao2018noise, hou2019domain}, the DAT approach was used to adapt SE models to new noisy conditions.

Despite yielding promising performance, the existing unsupervised domain adaptation approaches require additional information, such as word labels, language models, and noise-type labels. In this paper, we propose a new approach: discriminator-constrained OT network (DOTN) to perform unsupervised domain adaptation on SE. In contrast to related works, DOTN does not require additional label information when adapting the original SE models to match new noisy conditions. Our experiments show that DOTN can effectively adapt the SE models to new testing conditions, and that it achieves better adaptation performance than previous adversarial domain adaptation methods, which require additional noise type information.

\section{Method}
\subsection{Problem setting and notation }
Consider a source domain with paired data $(\mathbf{X}^s, \mathbf{Y}^s)= \left\{(\mathbf{x}^s_i, \mathbf{y}^s_i ) \right\}_{i=1}^{N_s}$, where $\mathbf{x}^s_i \in \mathbb{R}^n$, $\mathbf{y}^s_i \in \mathbb{R}^m$ stand for the input and the corresponding label of sample $i$. The unsupervised domain adaptation assumes that another target domain exists containing only data unlabeled,  $\mathbf{X}^t=\left\{\mathbf{x}^t_i \in \mathbb{R}^n \right\}_{i=1}^{N_s}$ . The goal is to seek a ground truth estimator (or statistical hypothesis) $f: \mathbb{R}^n \to \mathbb{R}^m$ for the target labels $\mathbf{Y}^t=\left\{\mathbf{y}^t_i\right\}_{i=1}^{N_t}$ (exist but not known), based on the knowledge provided by the source domain.

The probability distribution of a dataset $\mathcal{D}$ is denoted by $\mathbb{P}_{\mathcal{D}}$, where $\mathcal{D}$ is either $\mathbf{X}^s$, $\mathbf{Y}^s$, $\mathbf{X}^t$ or $\mathbf{Y}^t$ in our discussion. Our problem is to find a function $f$ such that $f$ induces a probability distribution $\mathbb{P}_{f(\mathbf{X}^t)}$ in $\mathbf{Y}^t$ with $\mathbb{P}_{f(\mathbf{X}^t)} \rightarrow \mathbb{P}_{\mathbf{Y}^t}$ under certain measure. We propose using the concept of OT to solve this problem.

\subsection{Proposed model: Discriminator-Constrained Optimal Transport Network (DOTN)}

Given a pair of distributions $\mathbb{P}_{\mathcal{D}_1}$ and $\mathbb{P}_{\mathcal{D}_2}$ and a displacement cost matrix $C \geq 0$, OT solves for the transportation plan $\gamma\in\prod(\mathbb{P}_{\mathcal{D}_1},\mathbb{P}_{\mathcal{D}_2})$ that minimizes the total cost (in a discrete setting)
\begin{equation}\label{E:total cost}
\min_{\gamma\in\prod(\mathbb{P}_{\mathcal{D}_1},\mathbb{P}_{\mathcal{D}_2})} \langle C, \gamma \rangle_F,
\end{equation}
where $\prod(\mathbb{P}_{\mathcal{D}_1},\mathbb{P}_{\mathcal{D}_2})$ denotes the space of joint distributions with marginal $\mathbb{P}_{\mathcal{D}_1}$ and $\mathbb{P}_{\mathcal{D}_1}$. $\langle \cdot ,\,\cdot \rangle_F$ is the Frobenius product, and the entry $C_{ij}$ of $C$ represents the displacement cost of the $i^{th}$ and $j^{th}$ samples. It can be proven that the minimum of this problem is a distance and called the Wasserstein distance when the corresponding cost is a norm~\cite{villani2008optimal, peyre2019computational}.

Our proposed method consists of two parts: OT alignment and Wasserstein Generative Adversarial Network (WGAN) training~\cite{arjovsky2017wasserstein, gulrajani2017improved}. Both steps are based on OT; however, they are considered with two different pairs of distributions and employ different algorithms.

\paragraph{Adaptation by Joint Distribution Optimal Transport}

Our adaptation mechanism relies on the alignment between the joint distributions of source and target domains (for the target domain, the label is estimated label). In particular, we approximate $f$ by minimizing the OT loss between the joint distributions $\mathbb{P}_{\mathbf{X}^s}\times \mathbb{P}_{\mathbf{Y}^s}$ and $\mathbb{P}_{\mathbf{X}^t}\times \mathbb{P}_{f(\mathbf{X}^t)} $, with a chosen cost matrix,
\begin{equation}\label{cost matrix}
C_{ij} = \alpha \, \| \mathbf{x}^s_i - \mathbf{x}^t_j \|^2 + \beta \, \| \mathbf{y}^s_i - f( \mathbf{x}^t_j ) \|^2, \quad \qquad  (\alpha, \beta > 0)
\end{equation}

By aligning the joint distributions of the source and target domains, noise adaptation is naturally achieved as OT seeks the source sample that is the most ``similar'' for each target sample.

Although the OT provides accurate estimates for each sample, the effect of each estimation error could accumulate in the training process and mislead $f$ to a convenient local minimum without preserving the speech data structure. To avoid this situation, we employ Wasserstein Generative Adversarial Network (WGAN) training to complement and enhance our adaptation system.

\paragraph{Discriminative training on outputs and source labels}

Distinct from the adaptation where we consider the joint distributions of the inputs and labels, we focus on WGAN training for the source label distribution $\mathbb{P}_{\mathbf{Y}^s}$ and output distribution $\mathbb{P}_{f(\mathbf{X}^t)}$. In the terminology of generative adversarial training, we consider $f$ a generator and introduce a convolutional neural network-based discriminator, $h$, as the 'critic'. In general, we use the discriminator to decide whether the outputs of $f$ are 'similar' to the source labels. Formally, the WGAN algorithm solves
\begin{equation}\label{Eq: WGAN}
\min_f\max_{h\in \mathcal{L}} \Big\{\mathbb{E}_{y\sim \mathbb{P}_{\mathbf{Y}^s}}(h(y)) - \mathbb{E}_{x\sim \mathbb{P}_{\mathbf{X}^t}}(h(f(x))) \Big\}
\end{equation}
by Kantorovich-Rubinstein duality~\cite{villani2008optimal}, where $\mathcal{L}$ is the set of 1-Lipschitz functions. In this case, under an optimal discriminator minimizing the value function with respect to the generator
parameters minimizes the Wasserstein distance between the distributions $\mathbb{P}_{\mathbf{Y}^s}$ and $\mathbb{P}_{f(\mathbf{X}^t)}$.

This discriminative training complements our alignment and provides additional constraints from the explicit relation between the source labels and estimations of target labels. These constraints support the joint distribution alignment and provides further guidance in the gradient descent training process. The performance of the experiments is considerably improved when our discriminative training supports the joint distribution alignment.

\subsection{Loss functions and the proposed algorithm}

Our domain alignment can be achieved by solving the following optimization problem:  
\begin{equation}\label{E:OT}
   \min_{\gamma, f} \mathcal{L}_1+\mathcal{L}_2=  \min_{\gamma, f}\frac{1}{N^s}\sum_{i} \|\mathbf{y}_i^s-f(\mathbf{x}_i^s)\|^2+\sum_{i,j}\gamma_{ij} \Big( \alpha \,  \|\mathbf{x}^s_i-\mathbf{x}^t_j\|^2 +\beta \, \|\mathbf{y}^s_i -f( \mathbf{x}^t_j)\|^2 \Big),
\end{equation}
where $\alpha,\beta>0$ are the parameters chosen for balance. Notably, the first term emphasizes the knowledge from the source domain is not to be forgotten during training, which has been revealed in several works ~\cite{damodaran2018deepjdot, li2017learning, shmelkov2017incremental}. Without this emphasis, the source domain knowledge cannot be well-maintained, and thus the overall performance may degrade. Such phenomenon was also observed in the SE experiments. The second term is intended for domain alignment. To show some intuitions, consider the ideal case where Eq.~(\ref{E:OT}) is completely minimized to zero, which leads to
\begin{equation}
\|\mathbf{x}^s_i - \mathbf{x}^t_j \|^2 \equiv 0 \,\, \mbox{and} \,\, \| \mathbf{y}^s_i -f( \mathbf{x}^t_j) \|^2 \equiv 0 \quad \Rightarrow \quad \mathbf{x}^s_i = \mathbf{x}^t_j  \,\, \mbox{and} \,\, \mathbf{y}^s_i  = f( \mathbf{x}^t_j)
\end{equation}
for all $i, j$. This indicates that for each given target domain sample $\mathbf{x}^t_j$, an identical sample $\mathbf{x}^s_i$ from the source domain is found, and the unknown target label is then constructed by the corresponding source label.
Even though practically the ideal case of zero loss is unlikely to happen, the OT loss aims to search for the most ``similar'' correspondence, which entails the intuition of domain alignment for Eq.~(\ref{E:OT}).  From this point of view, it is noted that although the term $\|\mathbf{x}^s_i-\mathbf{x}^t_j\|$ in $C_{ij}$ (in Eq. (\ref{cost matrix})) is not directly related to the network backpropagations of $f$ and $h$, it may not be ignored as the discard of this term will result in a wrong transportation plan $\gamma_{ij}$ and eventually lead to an undesired alignment.

For discriminative training, the discriminator $h$ is trained by the discriminator loss function $\mathcal{L}_h=\frac{1}{m}\sum_{i=1}^m h(\mathbf{y}_i^s)-h(f(\mathbf{x}^t_i)),$ and $f$ follows the generator loss function $\mathcal{L}_f= - \frac{1}{m}\sum_{i=1}^m h(f(\mathbf{x}^t_i))$ where $m$ is the batch size. As there are multiple sets of parameters $\gamma$, $h$ and $f$ in our framework, one set of parameters is updated each time, while the other sets of parameters are fixed. 
 
\begin{figure}
  \centering
  \includegraphics[width=0.68 \columnwidth]{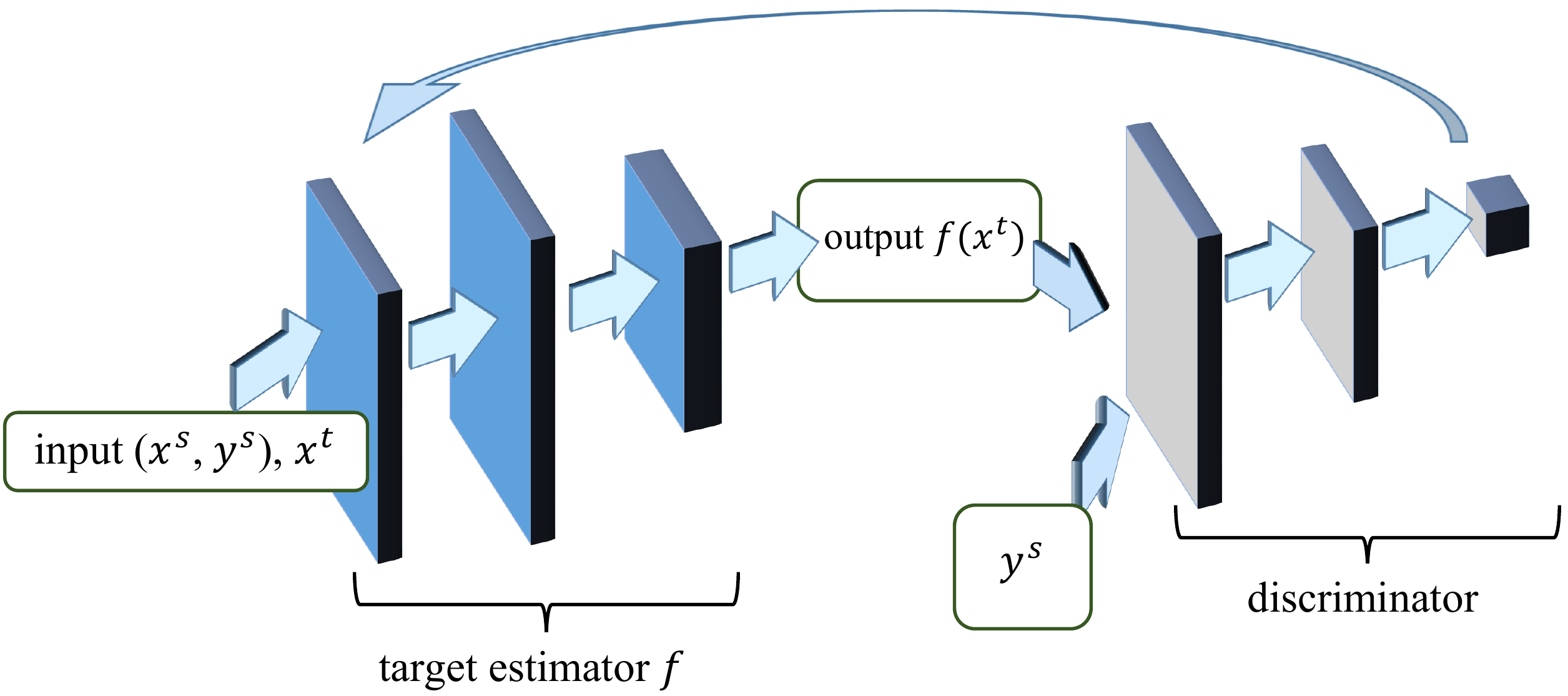}
  \caption{DOTN network structure.}
\end{figure}

\begin{algorithm}
\caption{DOTN, proposed algorithm}
\begin{algorithmic}[1]
\Require $\mathbf{x}^s$, source domain inputs. $\mathbf{y}^s$, source domain labels. $\mathbf{x}^t$, target domain inputs. $c$, the clipping parameter. $m$, the batch size. $n_f, n_h, n_s$: number of iterations of OT per generator training, discriminator training, and source domain training, respectively. $n$, number of iterations.
\Require $\theta_f$, initial parameters of estimator $f$. $\theta_h$, initial parameters of discriminator $h$.
\For{each batch of source samples $(\mathbf{x}^s, \mathbf{y}^s)$ and target samples $(\mathbf{y}^t)$}
\State fix $\theta_f$, solve for $\gamma$ in Eq.~(\ref{E:OT}) by OT.
\State fix $\gamma$, $\theta_f\leftarrow $Adam$(\nabla_{\theta_f}\mathcal{L}_2,\theta_f, \theta_h)$.
\If{$n$ mod $n_{s}==0$ }
\State $\theta_f\leftarrow$Adam$( \nabla_{\theta_f}\mathcal{L}_1, \theta_f, \theta_h)$.
\EndIf
\If{$n$ mod $n_{f}==0$ }
\State $\theta_f\leftarrow$Adam$( \nabla_{\theta_f}\mathcal{L}_f, \theta_f, \theta_h)$.
\EndIf
\If{$n$ mod $n_{h}==0$ }
\State $\theta_h\leftarrow $Adam$(\nabla_{\theta_h}\mathcal{L}_h, \theta_f,  \theta_h)$.
\State $\theta_h\leftarrow$clip$(\theta_h, -c,c).$
\EndIf
\EndFor
\end{algorithmic}
\end{algorithm}

\section{Experiments}
We evaluated our method in SE on two datasets: Voice Bank corpus~\cite{veaux2013voice} and TIMIT~\cite{garofolo1988getting}. Although there is abundant literature on unsupervised domain adaptation, very limited methods have been successfully applied to SE or regression problems. Closely relevant methods often require additional input structures or domain label. Unlike previous methods, our approach does \textbf{not} rely on additional data information. Nevertheless, we compared our results with two most relevant adversarial domain adaptation methods: DAT~\cite{liao2018noise} and MDAN~\cite{zhao2018adversarial}. The original MDAN was designed for classification problems and cannot be directly applied on SE. Preserving the fundamental ideas, we modified the MDAN structure for regressions, so that SE experiments can be performed for comparison.  The implementations are summarized in the supplementary material and codes are available on Github\footnote{\href{https://github.com/hsinyilin19/Discriminator-Constrained-Optimal-Transport-Network}{https://github.com/hsinyilin19/Discriminator-Constrained-Optimal-Transport-Network} }.

\subsection{Comparisons to DAT and MDAN}
As both DAT\footnote{DAT consists of three components, a deep feature extractor $\mathcal{E}: \mathcal{X} \to \mathcal{Z}$, a (task) label predictor $F_Y: \mathcal{Z} \to \mathcal{Y}$, and a domain classifier $F_D: \mathcal{Z} \to \mathcal{K}$, to form two functional pairs: $F_D \circ \mathcal{E}$ and $F_Y \circ \mathcal{E}$. Here $\mathcal{X}$, $\mathcal{Y}$ are the input space and (task) label space respectively. $\mathcal{Z}$ denotes the invariant (latent) feature space; $\mathcal{K} = \{0:\mbox{source}, 1:\mbox{target}\}$ as the domain label classes. The pair $F_D \circ \mathcal{E}: \mathcal{X} \to \mathcal{K}$ formed by $F_D$, $\mathcal{E}$ works against each other, by a Gradient Reversal Layer, to derive domain invariant features in $\mathcal{Z}$. Another pair $F_Y \circ \mathcal{E}: \mathcal{X} \to \mathcal{Y}$ demands that the invariant features in $\mathcal{Z}$ encode sufficient information for (main task) label classifications at the same time. Via adversarial training, the two pairs eventually reach a balance completing the main learning task as well as eliminating the domain mismatch.} and MDAN\footnote{MDAN inherits the DAT architecture with an extension to $K$ source domains and $K$ domain classifiers $F_{D_i}: \mathcal{Z} \to \{0, 1\}$, $i=1,\ldots K$ with similar adversarial training applied.} utilize a domain classifier to derive domain invariant features, domain labels become essential in their adaptation mechanism. Thus, both these methods are considered \textit{weakly supervised}. Specifically, they request inputs of the form $\left\{\mathbf{x}_i^s, \mathbf{y}^s_i, c_i\right\}_i$ and $\left\{\mathbf{x}_j^t, c_j\right\}_j$ with $c_i, c_j \in \mathcal{K}$, where $\mathcal{K}$ is an index set specifying the origin of data domains. For DAT, $\mathcal{K}=\{0, 1\}$ to indicate whether a sample belongs to the source or target domain, while for MDAN $\mathcal{K} = \{0, 1, \ldots, K\}$ to indicate that there are $K$ distinct source domains and one target domain. The requirement on domain labels poses restrictions in certain circumstances, such as a new target sample may not always fall into any existing categories, or the data origin is simply unknown. In contrast, the proposed method DOTN is not bounded by specifications of data origin and simply receives inputs of the form $\left\{\mathbf{x}_i^s, \mathbf{y}^s_i, \mathbf{x}_j^t\right\}_{i,j}$. With less input requirement, DOTN is more flexible to be applied in various scenarios, more approachable to real-world applications.

\subsection{Voice Bank with DEMAND noise database}

\paragraph{Dataset} In the first set of experiments, the pre-selected subset of Voice Bank provided by~\cite{valentini2016investigating} was used to test the proposed DOTN. For source domain data, 14 male speakers and 14 female speakers were randomly selected out of totally 84 speakers (42 male and 42 female), and each speaker pronounced around 400 sentences. As a result, the clean data in source domain contained 5,724 male utterances and 5,848 female utterances, amounting to 11,572 utterances in total. We then mixed the 11,572 clean utterances with noise from DEMAND~\cite{thiemann_joachim_2013_1227121}, in transportation category: ``Bus'', ``Car'', ``Metro'' at 7 Signal-to-noise ratio (SNR) levels (-9, -6, -3, 0, 3, 6, and 9 dB). Accordingly, for this set of source domain data, both noisy speech signals and the corresponding clean references are prepared.  

A target domain data contained $5,768$ noisy utterances mixed by 824 clean ones from two random speakers (1 male, 1 female) (followed the design in ~\cite{pascual2017segan}) with one of the three noise types from DEMAND, in STREET category: ``Traffic'', ``Cafe'', or ``Public square'' and 7 SNR ratios (-9, -6, -3, 0, 3, 6, and 9 dB). No clean labels were given under target domain. That is, for the target domain data, only noisy speech signals are provided and the corresponding clean references are not accessible. 

All source samples (both noisy utterances and the corresponding clean references, prepared from the three transportation noise types) and the target samples (only noisy utterances without the corresponding clean references, prepared from the one street noise type) were included in our training set. We conducted the experiments under single-target noise-type circumstances and compared the results with DAT and MDAN. More specifically, we ran this setting for all three cases, where the target domain contained either cafe, public square, or traffic noise.

\paragraph{Results}
Table~\ref{VBD_table_pesq} and Table~\ref{VBD_table_stoi} list the PESQ and STOI scores, respectively, of DAT, MDAN and DOTN under three noise types at 7 SNR levels. ``Avg'' denotes the averaged scores over 7 SNR levels. From the PESQ scores reported in Table~\ref{VBD_table_pesq}, we note that the proposed DOTN outperforms both DAT and MDAN consistently over different noise types and SNR levels, except for the Cafe noise type at -9 dB SNR. This might be owning to a potential limitation of DOTN, which will be detailed in the next section. Next, from Table~\ref{VBD_table_stoi}, we note that STOI scores show very similar trends to that of PESQ scores, as listed in Table~\ref{VBD_table_pesq}.

\begin{table}
\caption{PESQ scores for VoiceBank-DEMAND} 
\label{VBD_table_pesq}
\centering 
\resizebox{\columnwidth}{!}{%
\begin{tabular}{cccccccccc} 
\toprule

 noise type  & \multicolumn{3}{c}{\textbf{Traffic}}& \multicolumn{3}{c}{\textbf{Cafe}}& \multicolumn{3}{c}{\textbf{Public square}} \\ [0.5ex]
\cmidrule(r){2-10}
 SNR(dB)/model & \textbf{DAT}~\cite{liao2018noise} &\textbf{MDAN}~\cite{zhao2018adversarial} &\textbf{DOTN} &\textbf{DAT} &\textbf{MDAN} & \textbf{DOTN}  &\textbf{DAT} &\textbf{MDAN}  &\textbf{DOTN}\\

 \cmidrule(r){1-10}
-9  &1.307  &1.670    &\textbf{1.863}   &1.058  &\textbf{1.539} &1.497            &1.436  &1.929   &\textbf{2.037} \\
-6  &1.446  &1.920    &\textbf{2.182}   &1.184  &1.735          &\textbf{1.853}   &1.655  &2.119   &\textbf{2.258} \\
-3  &1.718  &2.153    &\textbf{2.395}   &1.362  &1.949          &\textbf{2.089}   &1.939  &2.318   &\textbf{2.439} \\
0   &2.081  &2.366    &\textbf{2.591}   &1.599  &2.153          &\textbf{2.332}   &2.268  &2.512   &\textbf{2.601} \\
3   &2.381  &2.535    &\textbf{2.740}   &1.865  &2.345          &\textbf{2.490}   &2.575  &2.670   &\textbf{2.761} \\
6   &2.712  &2.708    &\textbf{2.888}   &2.189  &2.534          &\textbf{2.661}   &2.867  &2.824   &\textbf{2.889} \\
9   &3.016  &2.854    &\textbf{3.015}   &2.493  &2.695          &\textbf{2.783}   &3.120  &2.966   &\textbf{3.057} \\

Avg &2.094 &2.315     &\textbf{2.525}   &1.679  &2.136          &\textbf{2.244}   &2.266  &2.477   &\textbf{2.577} \\
  
\bottomrule
\end{tabular}}
\end{table}

\begin{table}
\caption{STOI scores for VoiceBank-DEMAND} 
\label{VBD_table_stoi}
\centering 
\resizebox{\columnwidth}{!}{%
\begin{tabular}{cccccccccccc} 
\toprule

 noise type & \multicolumn{3}{c}{\textbf{Traffic}}& \multicolumn{3}{c}{\textbf{Cafe}}& \multicolumn{3}{c}{\textbf{Public square}} \\ [0.5ex]
 \cmidrule(r){2-10}

   SNR(dB)/model &\multicolumn{1}{c}{ \textbf{DAT}~\cite{liao2018noise}}&\multicolumn{1}{c}{ \textbf{MDAN}~\cite{zhao2018adversarial}}&\multicolumn{1}{c}{ \textbf{DOTN}}&\multicolumn{1}{c}{ \textbf{DAT} }&\multicolumn{1}{c}{ \textbf{MDAN} }&\multicolumn{1}{c}{ \textbf{DOTN}}&\multicolumn{1}{c}{ \textbf{DAT} }&\multicolumn{1}{c}{ \textbf{MDAN}}&\multicolumn{1}{c}{ \textbf{DOTN}}\\
\cmidrule(r){1-10}
-9  &0.584  &0.708   &\textbf{0.721}   &0.557  &\textbf{0.643}   &0.633           &0.667 &0.747  &\textbf{0.765} \\
-6  &0.659  &0.761   &\textbf{0.790}   &0.616  &0.702            &\textbf{0.723}  &0.725 &0.792  &\textbf{0.815} \\
-3  &0.728  &0.810   &\textbf{0.833}   &0.687  &0.760            &\textbf{0.779}  &0.771 &0.830  &\textbf{0.850} \\
0   &0.785  &0.849   &\textbf{0.871}   &0.741  &0.805            &\textbf{0.831}  &0.815 &0.859  &\textbf{0.877} \\
3   &0.824  &0.874   &\textbf{0.894}   &0.785  &0.842            &\textbf{0.861}  &0.845 &0.881  &\textbf{0.900} \\
6   &0.860  &0.896   &\textbf{0.914}   &0.825  &0.871            &\textbf{0.887}  &0.872 &0.898  &\textbf{0.915} \\
9   &0.885  &0.910   &\textbf{0.927}   &0.854  &0.893            &\textbf{0.905}  &0.893 &0.914  &\textbf{0.931} \\

Avg  &0.761 &0.830   &\textbf{0.850}   &0.724  &0.788            &\textbf{0.803}  &0.798 &0.846  &\textbf{0.865} \\
  
\bottomrule
\end{tabular}}
\end{table}

\subsection{TIMIT}

\paragraph{Dataset}
For the second part of experiments, TIMIT was used to prepare the source and target samples. The \textit{clean speech} of the source domain $\{ \mathbf{y}^s_i \}_{i=1}^{N_s}$ for training consists of utterances, contributed by 48 male speakers and 24 female speakers from 8 dialect regions. Each speaker had 8 sentences, including 5 SX (phonetically compact sentences) and 3 SI (phonetically diverse sentences), according to the official suggestion of TIMIT. The number of the speakers selected was to maintain the balance of the original data.

The above clean utterances were used to mix with 5 \textit{stationary} noise types (Car, Engine, Pink, Wind, and  Cabin) at 9 SNR levels (-12, -9, -6, -3, 0, 3, 6, 9, and 12 dB), amounting to 25,920 noisy utterances, to be the \textit{noisy speech} of the source domain $\{\mathbf{x}^s_i\}_{i=1}^{N_s}$ with $N_s = 25,920$.

For the target domain, a total of 24 speakers suggested by TIMIT core test set was all used to have 192 clean utterances for the target domain $\mathbf{y}^t_i$, which were subsequently mingled with one of the four \textit{non-stationary} noise types: ``Helicopter'', ``Cafeteria'', ``Baby-cry'', or ``Crowd-party'' under 7 SNRs (-9, -6, -3, 0, 3, 6, and 9 dB) as target inputs $\{\mathbf{x}^t_i\}_{i=1}^{N_t}$ with $N_t = 1,344$. The choice of noise types for the source and target domain was to let the learning algorithms adapt from distinguished environments in the real-world.

\paragraph{Results}
Table~\ref{TIMIT_table} lists the PESQ and STOI scores of the DAT, MDAN, and DOTN under four noise types at seven SNR levels. From the table, we first note that DOTN consistently outperforms DAT and MDAN in terms of both ``Avg'' PESQ and STOI scores among the four noise types. With a more careful investigation, the DOTN achieves higher PESQ and STOI scores over DAT and MDAN for all SNR conditions in the Helicopter noise. For the Crowd-party and Cafeteria noises, DOTN outperforms DAT and MDAN in most higher SNR conditions for both PESQ and STOI scores. However, for the Baby-cry noise, DOTN outperforms DAT and MDAN in STOI but underperforms MDAN in PESQ. Note that Cafeteria, Crowd-party, and Baby-cry noise types involved clear human speech components, which may cause confusions when DOTN tries to retrieve the target speech (clean reference). Thus, DOTN yields sub-optimal performance when dealing with these noise types, especially under very low SNR conditions, where background speech components might overwhelm the target speech. Nevertheless, the overall average PESQ and STOI scores of DOTN (P=1.838; Q=0.7238) are still higher than that of DAT (P=1.5728; Q=0.6298) and MDAN (P=1.8001; Q=0.7038), where P and Q denote the PESQ and STOI scores, respectively, over the 4 noise types and 9 SNR levels. Please also note that DAT and MDAN require additional domain label information~\cite{liao2018noise} while DOTN performs domain adaptation in a purely unsupervised manner.

\begin{figure}
  \centering
  \includegraphics[width=0.9\columnwidth]{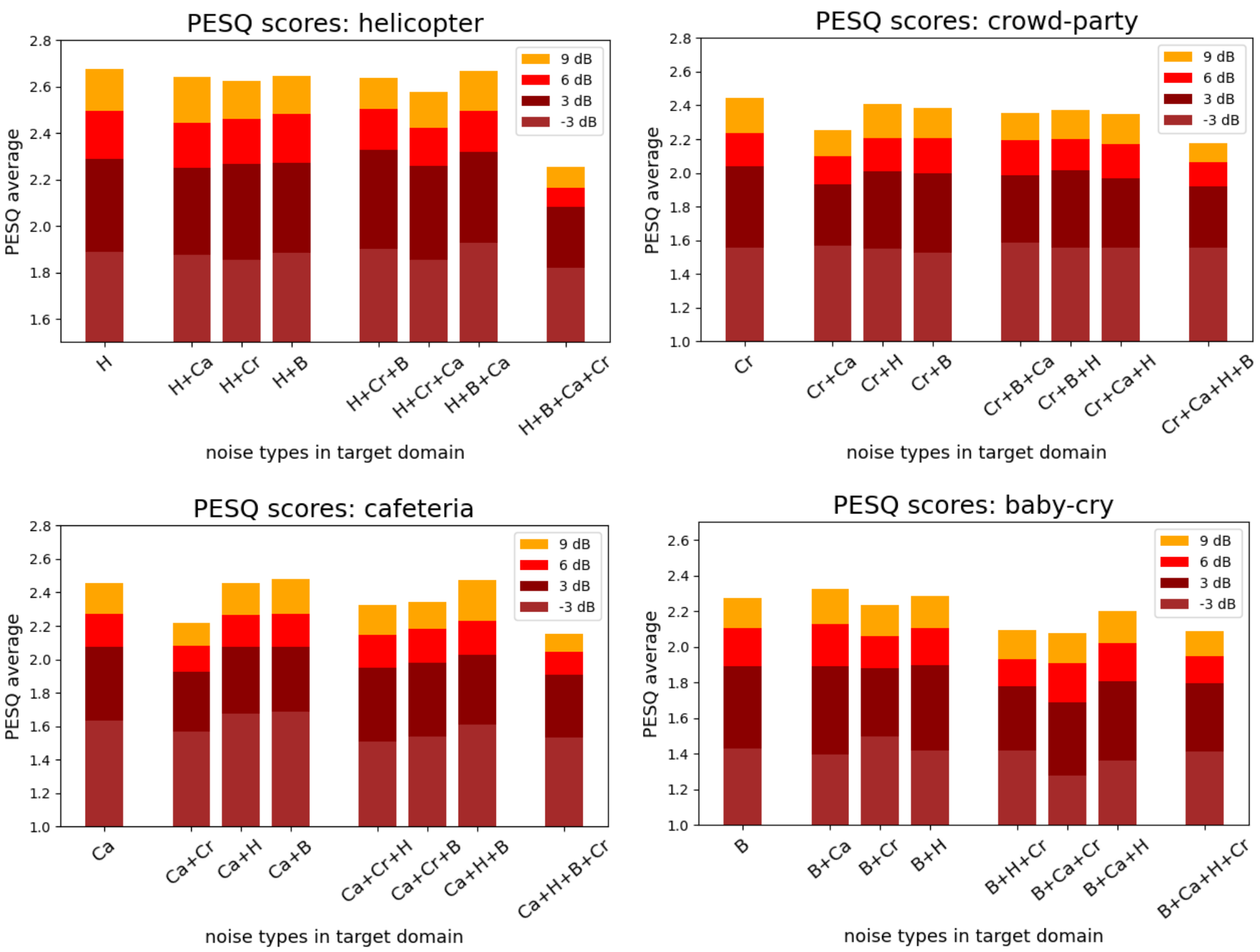}
  \caption{Comparisons in PESQ average for cases with multiple noise types in target domain, where H, Ca, Cr, and B denote Helicopter, Cafeteria, Crowd-party, and Baby-cry noise types.}
  \label{TIMIT_PESQ}
\end{figure}

\begin{figure}
  \centering
  \includegraphics[width=0.9\columnwidth]{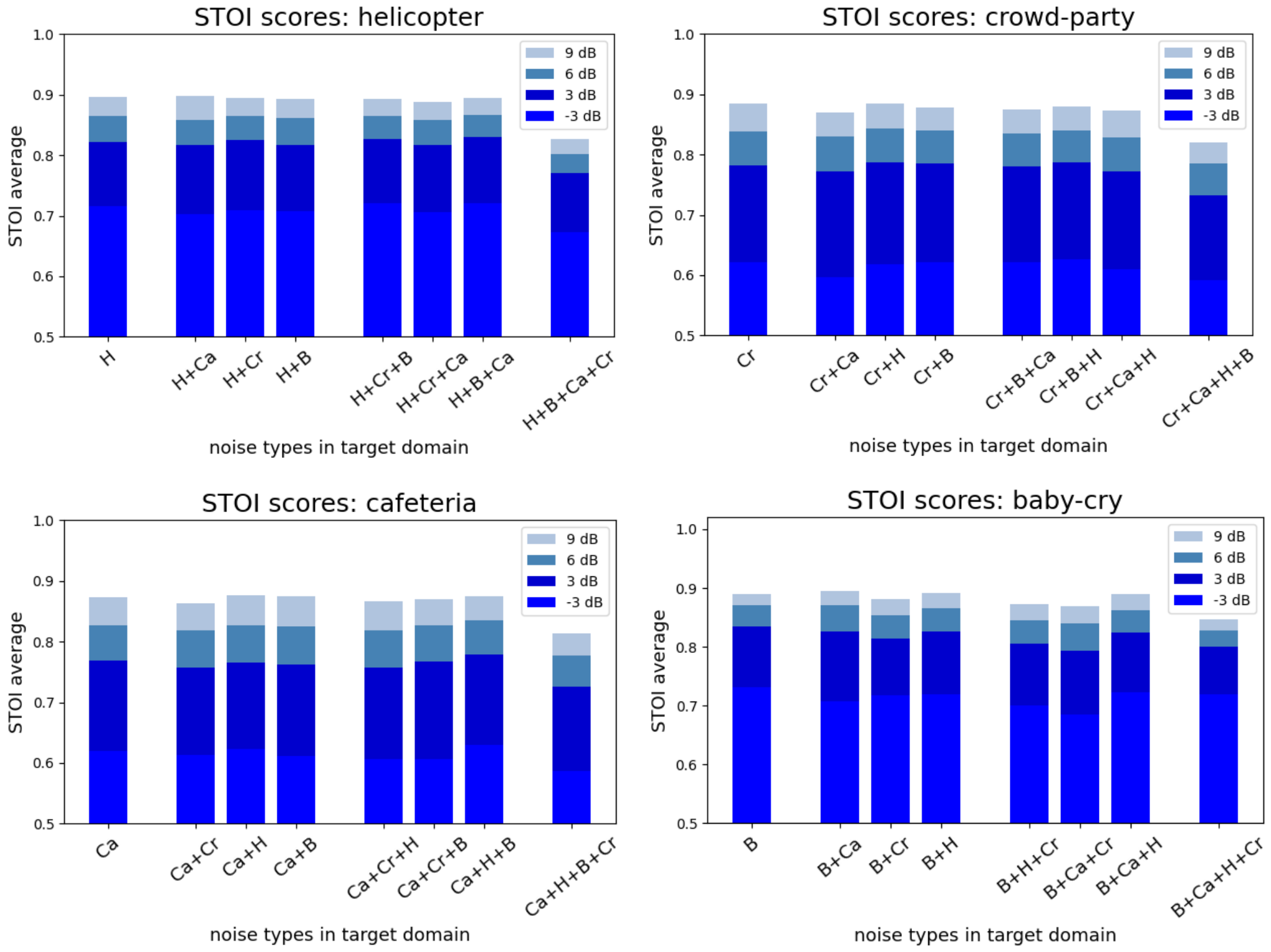}
  \caption{Comparisons in STOI average for cases with multiple noise types in target domain.}
  \label{TIMIT_STOI}
\end{figure}

To show the advantages and flexibility of DOTN, we further explored and applied DOTN to the circumstances when multiple noise types are included in the target domain. Fig~\ref{TIMIT_PESQ} and Fig~\ref{TIMIT_STOI}, respectively, present the PESQ and STOI results over four SNR levels on multiple target noise types; here H, Ca, Cr, and B denote Helicopter, Cafeteria, Crowd-party, and Baby-cry noise types, respectively. In Fig~\ref{TIMIT_PESQ}, when the testing noise is ``Helicopter'', the PESQ scores of using two noise types (H+Ca, H+Cr, H+B) and three noise types (H+Cr+B, H+Cr+Ca, H+B+Ca) are comparable. Further, the PESQ scores of these six systems are also comparable to that of using one noise type (H). The results confirmed that the DOTN has minor effects on catastrophic forgetting, which is a common issue in domain adaptation approaches~\cite{choy2006neural, goodfellow2013empirical}. In the case of 4 noise types (the system sequentially learned Helicopter, Baby-cry, Cafeteria, and Crowd-party noise types and was tested on the Helicopter noise), the achieved performance drops moderately. We observe very similar trends for the other three noise types (baby-cry, cafeteria, and crowd-party) in Fig~\ref{TIMIT_PESQ}. 

From Fig~\ref{TIMIT_STOI}, similar trends as those from Fig~\ref{TIMIT_PESQ} are observed: (1) The seven results using 1, 2, and 3 noise types are comparable, thereby confirming that the DOTN has minor effects on catastrophic forgetting. (2) When there are 4 noise types for sequential learning, the achievable STOI scores start to decrease.

\begin{table}
\caption{TIMIT results} 
\label{TIMIT_table}
\centering 
\resizebox{\columnwidth}{!}{%
\begin{tabular}{ccccccccccccc} 
\toprule
noise type &\multicolumn{6}{c}{\textbf{Helicopter}}&\multicolumn{6}{c}{ \textbf{Crowd-party}}\\
\cmidrule(r){2-13}
model &\multicolumn{2}{c}{\textbf{DAT}~\cite{liao2018noise}}&\multicolumn{2}{c}{\textbf{MDAN}~\cite{zhao2018adversarial} }&\multicolumn{2}{c}{\textbf{DOTN}}&\multicolumn{2}{c}{ \textbf{DAT} }&\multicolumn{2}{c}{\textbf{MDAN} }&\multicolumn{2}{c}{\textbf{DOTN}}\\
\cmidrule(r){2-13}
  SNR(dB) & PESQ & STOI & PESQ & STOI
 & PESQ & STOI & PESQ & STOI & PESQ & STOI & PESQ & STOI\\ [0.5ex]
 \cmidrule(r){1-13}
-9  &1.031  &0.392   &1.252  & 0.517 &\textbf{1.455}  &\textbf{0.577}  &\textbf{1.483}  &\textbf{0.544} & 1.150  &0.440   &1.056  &0.451\\
-6  &1.015  &0.431   &1.443  &0.594  &\textbf{1.669}  &\textbf{0.649}  &\textbf{1.484}  &\textbf{0.560} &1.356   &0.516   &1.302  &0.538\\
-3  &1.094  &0.497   &1.664  &0.670  &\textbf{1.890}  &\textbf{0.716}  &1.528           &0.592          &\textbf{1.560}   &0.600  &1.559 & \textbf{0.621}\\
0   &1.268  &0.566   &1.902  &0.742  &\textbf{2.104}  &\textbf{0.775}  &1.596           &0.636          &1.776            &0.684  &\textbf{1.816}  & \textbf{0.709}\\
3   &1.518  &0.637   &2.134  &0.801  &\textbf{2.289}  &\textbf{0.822}  &1.736           &0.690          &1.986            &0.762  &\textbf{2.042} & \textbf{0.782}\\
6   &1.779  &0.701   &2.363  &0.849  &\textbf{2.497}  &\textbf{0.865}  &1.953           &0.750          &2.179            &0.823  &\textbf{2.236} & \textbf{0.838}\\
9   &2.094  &0.759   &2.563  &0.884  &\textbf{2.677}  &\textbf{0.895}  &2.200           &0.801          &2.355            &0.865  &\textbf{2.447} &\textbf{0.885} \\

Avg  &1.400 &0.569   & 1.903& 0.722& \textbf{2.083} & \textbf{0.757}   &1.711          &0.653           &1.766            &0.670  &\textbf{1.780} &\textbf{0.690} \\
\cmidrule(r){1-13}

noise type &\multicolumn{6}{c}{ \textbf{Cafeteria}}&\multicolumn{6}{c}{\textbf{Baby-cry}}\\
\cmidrule(r){2-13}
model  &\multicolumn{2}{c}{\textbf{DAT}}&\multicolumn{2}{c}{\textbf{MDAN}} &\multicolumn{2}{c}{\textbf{DOTN}}&\multicolumn{2}{c}{\textbf{DAT}}&\multicolumn{2}{c}{\textbf{MDAN}}&\multicolumn{2}{c}{\textbf{DOTN}}\\
\cmidrule(r){2-13}
  SNR(dB) & PESQ & STOI & PESQ & STOI
 & PESQ & STOI & PESQ & STOI & PESQ & STOI & PESQ & STOI\\ [0.5ex]
\hline 
-9  &1.196 & 0.440  & \textbf{1.248}& \textbf{0.471}& 1.185 &0.436 & 1.109 & 0.580 &\textbf{1.116}& 0.561& 0.984 & \textbf{0.597}\\
-6  &1.206 & 0.471  &1.395& \textbf{0.535}&  \textbf{1.419}&  0.528 &\textbf{1.313}&  0.630 &\textbf{1.313}& 0.636& 1.180 & \textbf{0.668}\\
-3  &1.244 & 0.519  &1.608& 0.614&  \textbf{1.631} &\textbf{0.620}  &1.495 & 0.665 &\textbf{1.500}& 0.699& 1.431&  \textbf{0.732}\\
0  &1.408 & 0.579   &1.827& 0.692& \textbf{1.859}& \textbf{ 0.699}  & 1.621&  0.707 &\textbf{1.734}& 0.768& 1.665 & \textbf{0.789}\\
3  &1.631 & 0.651   &2.031& 0.764& \textbf{2.075} & \textbf{0.769} & 1.801 & 0.746 & \textbf{1.909}& 0.809& 1.889 & \textbf{0.835}\\
6  &1.915 & 0.719 & 2.244& 0.822&   \textbf{2.271} & \textbf{0.826}  &1.973 &  0.782 &\textbf{2.112}& 0.849& 2.105 & \textbf{0.871}\\
9  &2.216 & 0.782  & 2.422& 0.864&  \textbf{2.458} &\textbf{0.873}   & 2.133 & 0.814  &2.274& 0.880& \textbf{2.277} & \textbf{0.890} \\
 Avg  &1.545 &0.594 & 1.825& \textbf{0.680}&   \textbf{1.843} & 0.679   & 1.635&  0.703 &\textbf{1.708}&  0.743& 1.647 &\textbf{0.769} \\
\bottomrule
\end{tabular}}
\end{table}

\section{Conclusion}
In this study, we proposed a novel DOTN method, which was designed specifically for unsupervised domain adaptation in regression setting. Our approach skillfully fuses OT and generative adversarial frameworks to achieve unsupervised learning in target domain based on the information provided from the source domain, requiring no additional structure or inputs, such as multi-source domain and noise type labels. Our experiments show that the proposed method is capable of superior adaptation performance in SE by outperforming other adversarial domain adaptation methods on both PESQ and STOI scores for the VoiceBank-DEMAND and TIMIT datasets. Further, we show that when moderately increasing the complexity of the target samples (by increasing the number of noise types in the target domain), only a small degree of degradation was observed. This suggests that our method is robust to sample complexity in the target domain.

\bibliographystyle{unsrtnat}
\bibliography{main}



\section*{Supplementary material}

# Discriminator-Constrained Optimal Transport for Unsupervised Noise Adaptive Speech Enhancement –Supplementary Material–

**Hsin-Yi Lin**
The Cooperative Institute for Research in Environmental Sciences (CIRES)
University of Colorado, Boulder, CO, 80309, USA
NOAA Physical Sciences Laboratory, Boulder, CO, 80305, USA
`hylin@colorado.edu`

**Huan-Hsin Tseng**
Research Center for Information Technology Innovation
Academia Sinica, Taiwan
`htseng@citi.sinica.edu.tw`

**Xugang Lu**
National Institute of Information and Communications Technology (NICT), Japan
`xugang.lu@nict.go.jp`

**Yu Tsao**
Research Center for Information Technology Innovation
Academia Sinica, Taiwan
`yu.tsao@citi.sinica.edu.tw`

In this supplementary material, we provide implementation details, audio demonstrations, enhancement quality evaluations, as well as spectrograms and waveforms of some experimental results. The corresponding audio files of the presented results are accessible at `https://drive.google.com/drive/folders/1fuZIqM-feg4CUnU-zNeUCl_6I0ARJ0Pg?usp=sharing` and the codes of proposed method in the Github repository: `https://github.com/hsinyilin19/Discriminator-Constrained-Optimal-Transport-Network`.

## 1 Implementation details

**Corpus** We used Voice Bank (VCTK) [1] and TIMIT datasets available online. VCTK can be downloaded at https://datashare.is.ed.ac.uk/handle/10283/3443 and TIMIT can be found at: https://catalog.ldc.upenn.edu/LDC93S1.

**Noise database** The environmental noise recordings- DEMAND [2] mixed with Voice Bank in the experiments can be downloaded at: `https://doi.org/10.5281/zenodo.1227121`. The five stationary noises (Car, Engine, Pink, Wind, and Cabin) and four nonstationary noises (Helicopter, Cafeteria, Baby-cry, and Crowd-party) used in the TIMIT experiment can be found in our Github repository.

**Data processing** All corpora are in the WAV format with a 16 kHz sampling rate. Data preprocessing code is provided to generate clean speech from scratch for training and testing stage. Additionally,

another code mixing selected types of noise at a variety of SNR levels to clean utterances is provided. For computational convenience, waveforms were converted into STFT spectrograms.

**Network structures**

- The DAT [3] model was based on the optimal architecture provided on Github: https://github.com/jerrygood0703/noise_adaptive_DAT_SE, where two consecutive Bi-directional Long Short-Term Memory (BiLSTM) of 512 hidden units were used to connect with one fully-connected-layer of 1024 nodes for the SE generator. The domain classifier consisted of one LSTM of 1024 hidden units to connect with a fully-connected-layer of 1024 nodes for binary classification.

- Since MDAN [4] was not designed for regression tasks in the first place, much modification was required to fit the SE purpose. The original design used $k$ label (task) classifiers as well as $k$ domain classifiers for each of $k$ source domains. First, an encoder consisted of one BiLSTM of 512 hidden units was used to encode the input (spectrum) into 512-dim domain-insensitive latent variables. Subsequently, the original $k$ label classifiers were replaced by $k$ SE generators for output dimension 257, each of which contained one BiLSTM of 512 hidden units and a fully-connected-layer of 1024 nodes for regression outputs. The original MDAN code of classifications is provided: https://github.com/hanzhaoml/MDAN.git; our modification for speech enhancement is https://github.com/hsinyilin19/Discriminator-Constrained-Optimal-Transport-Network.

  In the meantime, there were $k$ additional source domain classifiers attempting to produce 512-dim domain-insensitive latent variables by applying the technique of Gradient Reversal Layers; each source domain classifier was comprised of one LSTM of 1024 hidden nodes and one fully-connected-layer of 1024 nodes for final binary classification.

- For DOTN, the discriminator was comprised of two consecutive 2D-Convolutional Neural Network (CNN) of kernel size 5 and subsequently two fully-connected-layers (16384 nodes and 256 nodes) to discriminate signals from the generator or not (True/False), where ReLu was used in between layers and Sigmoid for the final output. The generator was composed of a 2-layer BiLSTM of 512 hidden units to connect with two fully-connected-layers of 1024 nodes and 512 nodes, respectively.

**Optimization and hyperparameters**

- DAT was based on Tensorflow 1.6, where the ADAM optimizer with learning rate $10^{-4}$ and batch size 16 was adopted to train the model with $10^5$ iterations for TIMIT and $5 \times 10^4$ iterations for VCTK, respectively.

- MDAN used the ADAM optimizer with learning rate $10^{-3}$ and batch size 1800 to train 60 epochs for TIMIT and 5 epochs for VoiceBank-DEMAND, respectively. The ratio between the two losses of SE generator and domain classifier is 0.001.

- DOTN used the ADAM optimizer for both discriminator and generator with batch size 1800 to train 10 epochs for VoiceBank-DEMAND and 60 epochs for TIMIT. There were several training steps in the proposed method, each was set at a different learning rate. The OT alignment was trained with learning rate $10^{-5}$, the source domain knowledge with $10^{-4}$, the generator training with $10^{-5}$, and discriminator training with $10^{-3}$. $\alpha$ and $\beta$ were both fixed at 1, and the clipping parameter for discriminator was set at $10^{-3}$.

  To balance OT and adversarial training, we used different training frequency for each part of the proposed method to reach different levels of control strengths. From our experience, a successful training commonly happens when the training frequency is set from high to low in the following order: OT alignment, discriminator training, and then generator training. For example, the discriminator training was performed once every 5 iterations of OT alignment, and generator training was once every 10 iterations of OT alignment for TIMIT. On VoiceBank-DEMAND, the OT alignment and discriminator training had the same frequency, but the generator training was performed once every 2 iterations of OT alignment.

**Hardware**   All experiments were run on one NVIDIA Tesla V100 GPU of 32 GB CUDA memory and 4 CPUs with 90 GB memory.

**Runtime**

- (DAT) Around 4 hours training time on TIMIT ($10^5$ training iterations), and 3 hours training time on VoiceBank-DEMAND ($5 \times 10^4$ training iterations)
- (MDAN) Approximately 4 hours on TIMIT (60 epochs), and approximately 30 hours on VoiceBank-DEMAND (5 epochs).
- (DOTN) It consumed approximately 5 hours for each DOTN trial on TIMIT (60 epochs), and approximately 15 hours on VoiceBank-DEMAND (10 epochs).

## 2 Additional experimental results

### 2.1 Visualization of SE outputs

In the main manuscript, we present and discuss the quantitative results (in-terms of PESQ and STOI scores) of the proposed DOTN and compared methods, namely, DAT and MDAN. In this supplementary file, we present the waveform and spectrogram plots of the enhanced utterances produced by MDAN, DAT, and DOTN for qualitative analyses. A spectrogram plot is a popular tool to analyze the time-frequency characteristics of speech signals [5]. In Figs. 1 and 2, respectively, we demonstrate the waveform and spectrogram plots for an utterance pronounced by a male speaker from Voice Bank (no.232) contaminated with Cafe background noise from DEMAND at 0dB SNR level; the corresponding clean reference and enhanced versions by MDAN, DAT, and DOTN are also presented. In both figures, the top panels present the noisy utterance (right) and its clean version (left). The bottom panels demonstrate the enhanced results provided by MDAN (left), DAT (middle), and DOTN (right). From Figs. 1 and 2, MDAN, DAT, and DOTN all successfully suppress noise components given the noisy utterance. Among them, DAT seems to yield the best noise suppression result. With a further investigation on the spectrogram plot (Fig. 2), however, we note that some detailed speech structures of the DAT output are distorted, and some speech components are removed, as marked by yellow rectangular regions. The results from Figs. 1 and 2 clearly show that the qualitative results are consistent with those of the quantitative results (PESQ and STOI scores) as reported in the main manuscript. Next, Figs. 3 and 4, respectively, demonstrate the waveform and spectrogram plots of an utterance pronounced by a female speaker in Voice Bank (no.257) contaminated with the Cafe background at 0 dB SNR level along with its clean reference and enhanced versions. From Figs. 3 and 4, we note the same trends as those from Figs. 1 and 2: (1) MDAN, DAT, and DOTN all effectively suppress noise components given the noisy input, and DAT seems to yield the best noise suppression result. (2) As compared to DAT, DOTN can more effectively preserve detailed speech structures, as marked by the yellow rectangular regions in Fig. 4.

We further drew the waveform and spectrogram plots of utterances pronounced by one male and one female speaker from the TIMIT dataset. Figs. 5 and 6, respectively, are the waveform and spectrogram plots for a male speaker (labeled MTLS0), and Figs. 7 and 8, respectively, are the waveform and spectrogram plots for a female speaker (labeled FDHC0); both utterances were contaminated with Cafeteria background at 0dB SNR level; the clean references and enhanced versions are also presented in the figures. The qualitative results of the TIMIT dataset show very similar trends to that of "VoiceBank+DEMAND" (from Figs. 1 to 4). All of the three methods MDAN, DAT, and DOTN can suppress noise components while DOTN provides better results than the other two methods. The advantages of DOTN over DAT are marked by yellow rectangular regions in the spectrogram plots (Figs. 6 and 8). In summary, from Figs. 1 to 8, we note that the qualitative results align well with the quantitative scores as reported in the main manuscript. Please also note that our listening tests indicate that enhanced utterances by DOTN yields better quality with less distortion effects as compared to MDAN and DAT. Please refer to our audio samples: https://drive.google.com/drive/folders/1fuZIqM-feg4CUnU-zNeUCl_6IOARJ0Pg?usp=sharing.

Finally, we would like to make remarks on the structure of the proposed method DOTN. While the mechanism for domain adaptation of DOTN relies mainly on the joint distribution OT, the adversarial training is crucial for the enhanced speech quality. In fact, the MSE loss of spectrum was involved in our OT alignment, which (if used solely) could result in 'impetuous' erasing effect on the speech data and low speech quality as we observed in experiments. This phenomenon was also observed in the case of DAT [3], also a MSE-based method. The introduction of discriminator is our solution for attacking this issue. Based on the clean utterances in source domain (as references), the discriminator

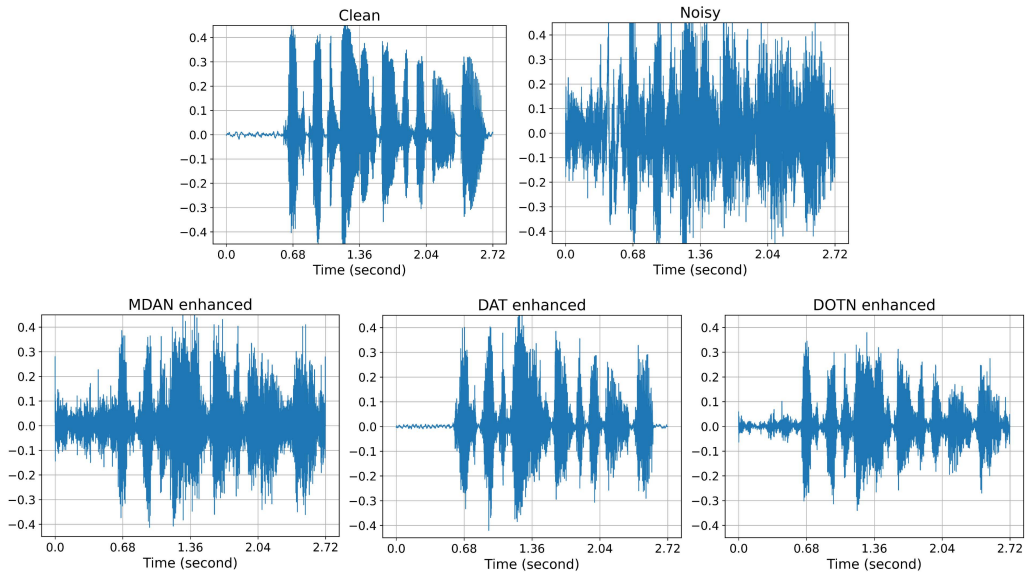

Figure 1: Waveforms for an utterance pronounced by a male speaker in Voice Bank (no.232) contaminated with Cafe background provided by DEMAND at SNR level 0 dB and its clean reference and enhanced versions.

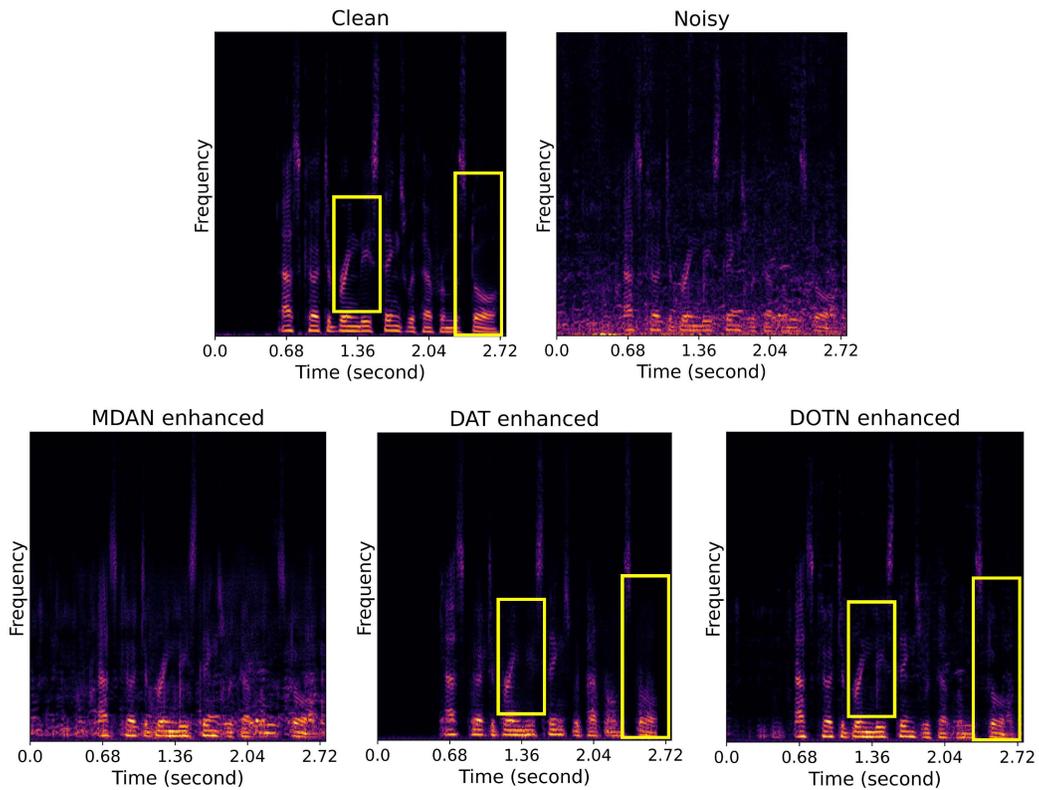

Figure 2: Spectrograms for an utterance pronounced by a male speaker in Voice Bank (no.232) contaminated with Cafe background provided by DEMAND at SNR level 0 dB and its clean reference and enhanced versions.

4

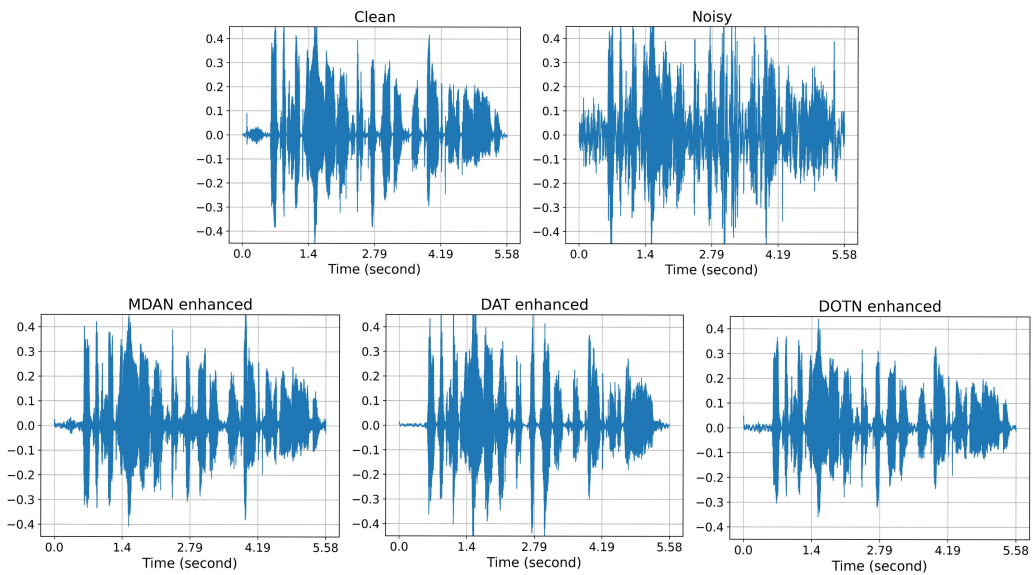

Figure 3: Waveforms for an utterance pronounced by a female speaker in Voice Bank (no.257) contaminated with Cafe background provided by DEMAND at SNR level 0 dB and its clean reference and enhanced versions.

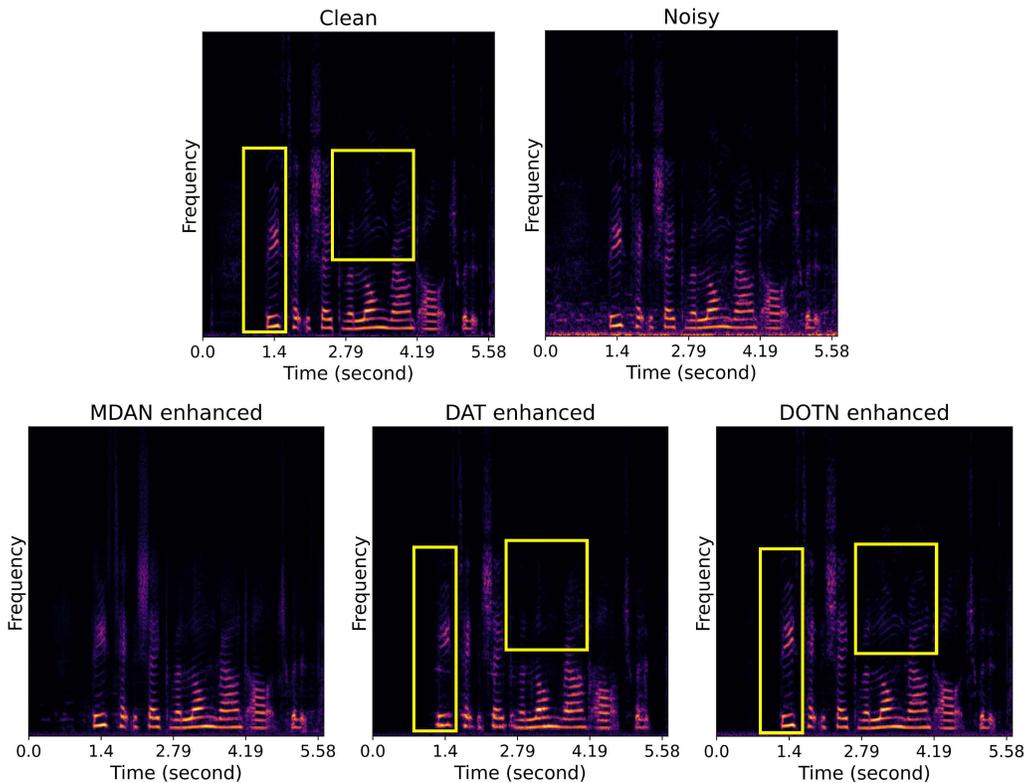

Figure 4: Spectrograms for an utterance pronounced by a female speaker in Voice Bank (no.257) contaminated with Cafe background provided by DEMAND at SNR level 0 dB and its clean reference and enhanced versions.

5

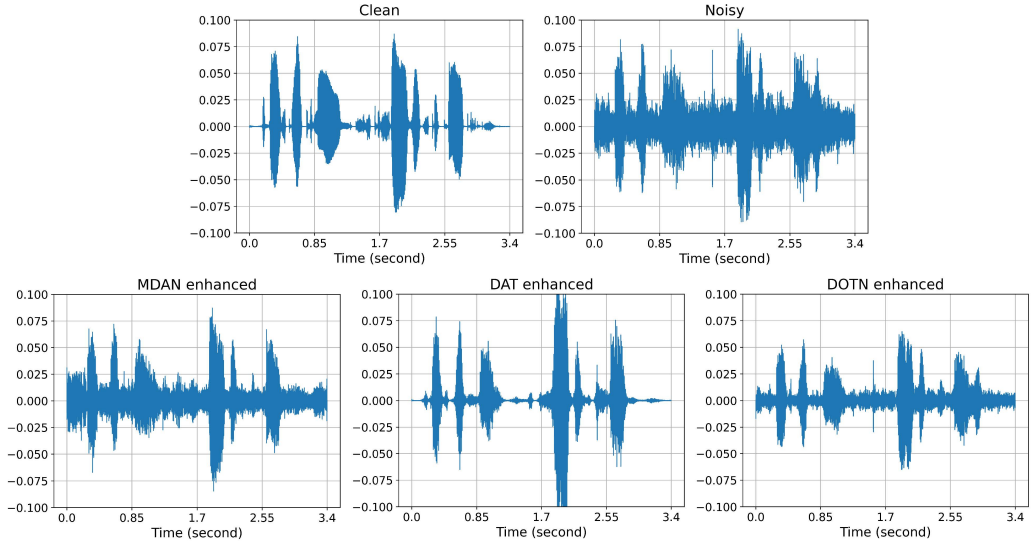

Figure 5: Waveforms for an utterance pronounced by a male speaker in TIMIT (labeled MTLS0) contaminated with Cafeteria background at SNR level 0 dB and its clean reference and enhanced versions.

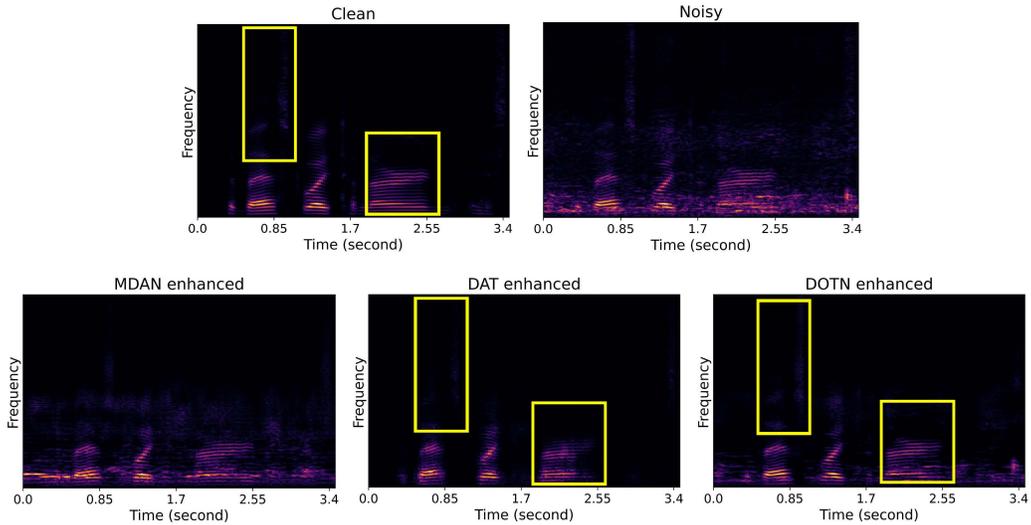

Figure 6: Spectrograms for an utterance pronounced by a male speaker in TIMIT (labeled MTLS0) contaminated with Cafeteria background at SNR level 0 dB and its clean reference and enhanced versions.

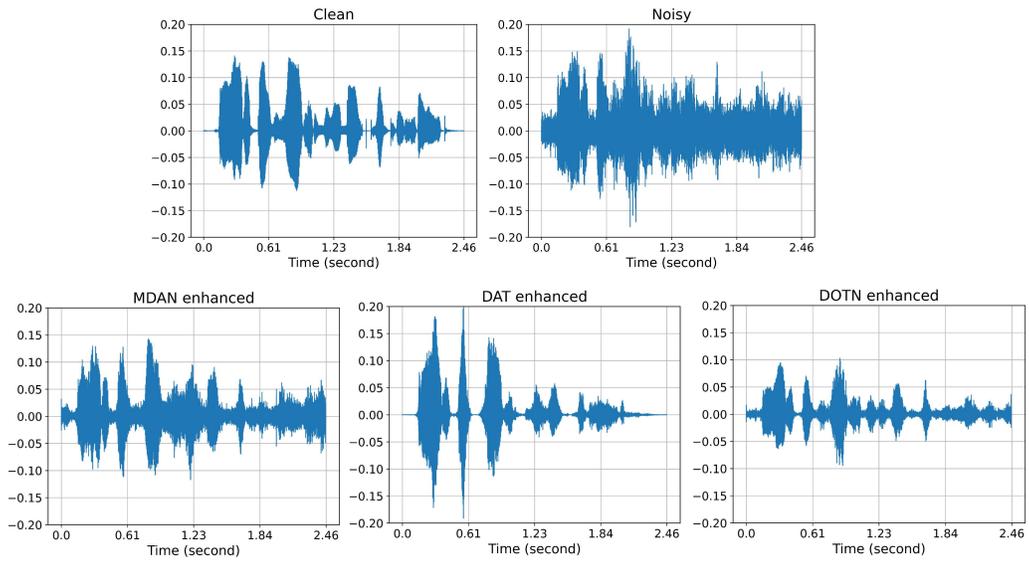

Figure 7: Waveforms for an utterance pronounced by a female speaker in TIMIT (labeled FDHC0) contaminated with Cafeteria background at SNR level 0 dB and its clean reference and enhanced versions.

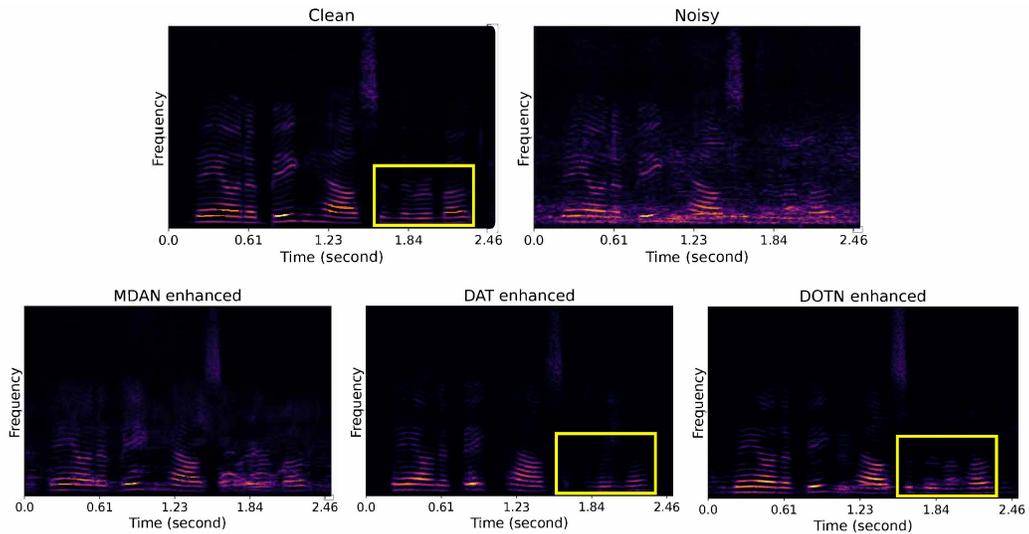

Figure 8: Spectrograms for an utterance pronounced by a female speaker in TIMIT (labeled FDHC0) contaminated with Cafeteria background at SNR level 0 dB and its clean reference and enhanced versions.

7

training renders a highly nonlinear constraint in the main OT alignment process to guarantee certain 'similarity' between the enhanced (fake) and clean (real) speech. One may question if it is appropriate to compare enhanced *target* data and clean *source* data. However, it is worth emphasizing that the task of discriminator is to roughly capture the character of natural clean speech, instead of making precise prediction in speech pattern. Due to the nature of this task, it does not post a logical issue when considering data in different domains.

## 2.2 Subjective evaluations

Subjective evaluations were conducted to collect individual opinions from their own perspectives, in contrast to formulated or objective metrics. The evaluation aimed to compare the proposed method with two weakly supervised method DAT and MDAN under human perceptions. 31 participants were gathered for blind test under random shuffles of audio recordings. All three enhancing methods appear in random orders; no knowledge of the audio source can be gained in advance. Participants were asked to rate each enhanced audio from 1 (bad) to 5 (excellent) for Mean Opinion Score (MOS) measurement. Only negative SNRs (dB) were used for demonstrating the significance of models, as well as reducing test time duration for participants.

For TIMIT, two denoised recordings were randomly chosen from 3 SNRs ($-3, -6, -9$), 4 target noises ("helicopter", "crowd-party", "cafeteria", "babycry") and 3 models (DAT, MDAN, and DOTN), which amounts to 72 audio recordings for TIMIT. Similarly in VCTK, two denoised recordings were randomly selected among 3 SNRs ($-3, -6, -9$), 3 target noises ("traffic", "CAFE", "public square") and 3 models (DAT, MDAN, DOTN), that amounts to 54 audio recordings. The MOS results of TIMIT and VCTK are listed Tables 1 and 2, respectively.

The results in Table 1, 2 showed that the human evaluations mostly favor DOTN over the other two methods. It is particularly dominant in the case of VCTK, Table 2, to confirm the perceptual performance of DOTN.

Table 1: MOS for TIMIT enhanced speech, ratings from 1 (bad) to 5 (excellent) for each audio recording given subjectively by each participant.

| noise type | **Helicopter** | | | **Crowd-party** | | |
|---|---|---|---|---|---|---|
| SNR/model | **DAT** | **MDAN** | **DOTN** | **DAT** | **MDAN** | **DOTN** |
| -9 dB | 2.08 | 2.03 | **2.16** | 1.84 | 1.47 | **2.13** |
| -6 dB | **2.76** | 2.24 | 2.66 | 1.89 | 1.90 | **2.68** |
| -3 dB | 3.10 | 2.32 | **3.11** | 2.68 | 2.24 | **3.10** |
| Avg | **2.65** | 2.20 | 2.64 | 2.14 | 1.87 | **2.64** |
| noise type | **Cafeteria** | | | **Babycry** | | |
| SNR/model | **DAT** | **MDAN** | **DOTN** | **DAT** | **MDAN** | **DOTN** |
| -9 dB | 1.84 | 1.47 | **2.61** | 2.71 | 1.58 | **2.71** |
| -6 dB | 2.24 | 1.61 | **2.74** | 3.26 | 2.37 | **3.47** |
| -3 dB | 2.56 | 2.19 | **3.35** | 3.40 | 2.60 | 3.37 |
| Avg | 2.21 | 1.76 | **2.90** | 3.12 | 2.18 | **3.18** |

Table 2: MOS for VCTK enhanced speech under the same setting as used in TIMIT.

| noise type | **Traffic** | | | **Cafe** | | | **Public square** | | |
|---|---|---|---|---|---|---|---|---|---|
| SNR/model | **DAT** | **MDAN** | **DOTN** | **DAT** | **MDAN** | **DOTN** | **DAT** | **MDAN** | **DOTN** |
| -9 dB | 2.10 | 2.05 | **3.68** | 2.15 | 1.76 | **3.89** | 3.05 | 2.44 | **3.85** |
| -6 dB | 2.56 | 2.81 | **4.11** | 2.35 | 2.19 | **3.61** | 2.74 | 2.52 | **4.00** |
| -3 dB | 3.19 | 2.76 | **4.23** | 3.40 | 2.47 | **4.34** | 3.76 | 3.29 | **4.29** |
| Avg | 2.62 | 2.54 | **4.01** | 2.63 | 2.14 | **3.95** | 3.18 | 2.75 | **4.05** |

### 2.3 Comparisons with the state-of-the-art supervised SE methods

Due to the lack of fully unsupervised domain adaptation methods for comparison with DOTN, an attempt to compare with the state-of-the-art (SOTA) "supervised" SE models may still be conducted to reveal some interests. A Transformer [6] model was used in the attempt of such comparison. A Transformer was trained from scratch on the datasets given in Sec. 4.2. Without domain adaptation, the Transformer as a supervised SE method was directly tested on the target domain. Table 3 showed the results with all SNRs summed over in three target domains.

It was observed that the DOTN had slightly better performance over the Transformer in most metrics. This result may be expected as the Transformer as a supervised SE method did not contain an adaptation mechanism to well adjusted to the target domain, in which case the background noise types were Cafe, Traffic, and Public Square.

Table 3: Comparisons to a state-of-the-art method in SE.

| noise/metric | Transformer PESQ/STOI | DOTN PESQ/STOI |
| --- | --- | --- |
| Cafe | 2.225/0.791 | **2.244/0.803** |
| Traffic | 2.496/0.840 | **2.525/0.850** |
| Public Square | **2.610**/0.858 | 2.577/**0.865** |

On the first glimpse, this does not seem a meaningful comparison, as the Transformer is supervised, while the proposed DOTN is designed for unsupervised adaptations. However, the comparison is based on the same training and testing sets, and the better performance of DOTN verifies the advantage of adaptation process. This advantage of adaptation should be more evident when we restrict the training set to be smaller. Another SOTA supervised SE method MetricGAN+ [7] can be used to conduct an additional comparison as the Transformer here. The full table and discussion can be found at https://drive.google.com/drive/folders/1cO3GCeFnQVXpatKXoyI0-b_PvfZwq6hO?usp=sharing.

### 2.4 An example on real-world applications

We also applied the adaptation models (DAT, MDAN, DOTN) to a real-world noisy speech data CHiME-3 for comparison; the enhanced audios can be found here: https://drive.google.com/drive/folders/1C4BiajlZfMjBXbemTtiA8PUIZpVhmQ4O?usp=sharing.

The audio demos consist of speech from two random male and female speakers selected from three noisy environments (Cafe, Street, Pedestrian). The DAT and MDAN models applied here on CHiME-3 were previously trained by adapting from the source noise domain: "Bus", "Car", "Metro" to target noise: "Traffic" under VCTK-DEMAND, as in the Sec. 4.2 of the paper.

As there is no clean speech for reference, DNSMOS [8] scores are computed alternatively for quality measures. The DNSMOS scores, provided here, averaging over audio samples are:

$$\text{noisy}: 2.73, \quad \text{DAT}: 3.12, \quad \text{MDAN}: 2.99, \quad \textbf{DOTN}: \textbf{3.27}$$

A possible reason for the degraded performance in DAT, MDAN compared to DOTN may be that the target domains appeared in CHiME-3: Cafe, Street, Pedestrian, did not appear in their pretrained target categories in the first place. *i.e.,* their domain classifiers had not seen such type of noise, and hence the domain mismatch may remain large.

On the other hand, the proposed DOTN aligns one domain to another by OT so that the data origin is not required. Many downstream tasks can therefore be achieved, especially when a new target domain without additional information on labels is confronted.

9



\end{document}